\documentclass[runningheads]{llncs}
\usepackage[T1]{fontenc}
\usepackage{subfig}
\usepackage{graphicx}
\usepackage{amsmath}
\usepackage{amssymb}
\usepackage{wrapfig}
\usepackage{verbatim}
\usepackage{upgreek}
\usepackage{booktabs}
\usepackage{multirow}
\usepackage{array}
\usepackage{listings}
\begin{document}
\title{Joint Graph Convolution for Analyzing\\Brain Structural and Functional Connectome}
\titlerunning{Joint GCN for Brain Connectome Analysis}
%
\author{
Yueting Li\inst{1} \and
Qingyue Wei\inst{1} \and
Ehsan Adeli\inst{1} \and
Kilian M. Pohl\inst{1,2} \and
Qingyu Zhao\inst{1}}
\authorrunning{Y. Li et al.}
\institute{Stanford University, Stanford, CA 94305, USA \and SRI International, Menlo Park, CA, 94025, USA\\}

\maketitle              
\begin{abstract} 
The white-matter (micro-)structural architecture of the brain promotes synchrony among neuronal populations, giving rise to richly patterned functional connections. A fundamental problem for systems neuroscience is determining the best way to relate structural and functional networks quantified by diffusion tensor imaging and resting-state functional MRI. As one of the state-of-the-art approaches for network analysis, graph convolutional networks (GCN) have been separately used to analyze functional and structural networks, but have not been applied to explore inter-network relationships. In this work, we propose to couple the two networks of an individual by adding inter-network edges between corresponding brain regions, so that the joint structure-function graph can be directly analyzed by a single GCN. The weights of inter-network edges are learnable, reflecting non-uniform structure-function coupling strength across the brain.  We apply our Joint-GCN to predict age and sex of 662 participants from the public dataset of the National Consortium on Alcohol and Neurodevelopment in Adolescence (NCANDA) based on their functional and micro-structural white-matter networks. Our results support that the proposed Joint-GCN outperforms existing multi-modal graph learning approaches for analyzing structural and functional networks. 
\end{abstract}
\section{Introduction}
The ``human connectom'' refers to the concept of describing the brain’s structural and functional organization as large-scale complex brain networks \cite{Sporns2005}. An accurate description of such connectome heavily relies on neuroimaging methods. Specifically, Diffusion Tensor Imaging (DTI) characterizes white-matter fiber bundles connecting different gray-matter regions (Structural Connectivity or SC), while resting-state functional Magnetic Resonance Imaging (rs-fMRI) measures spontaneous fluctuations in BOLD signal giving rise to the Functional Connectivity (FC) across brain regions \cite{moody2021}. Central to the connectome research is understanding how the underlying SC supports FC for developing high-order cognitive abilities \cite{Jung2016} and how the structure-function relationship develops over certain ages and differs between sexes \cite{baum2019}. 

To answer these questions, an emerging approach is using SC and FC of the brain to predict factors of interest, such as age, sex, and diagnosis labels \cite{Song2019isbi,Liu2019}. Since brain networks can be treated as graphs, one of the most popular architectures for building such prediction models is Graph Convolution Networks (GCN) \cite{kipf2016semi}. 
However, the majority of GCN-based studies on brain connectivity focus on a single imaging modality \cite{li2021braingnn,gadgil2020spatio,hanik2021predicting} but fall short in leveraging the relationship between FC and SC. Existing works that perform multi-modal fusion \cite{yalcin2021} either use two GCNs to extract features from SC and FC graphs separately \cite{zhang2018multi} or directly discard the graph structure from one modality (e.g., regarding FC as features defined on the SC graph) \cite{Liu2019}. These simple fusion methods ignore the inter-network dependency that white-matter fiber tracts  provide an anatomic foundation for high-level brain function, thereby possibly leading to sub-optimal feature extraction from the SC-FC profile.  

To address this issue, we propose coupling the SC and FC graphs by inserting inter-network edges between corresponding brain regions. The resulting joint graph then models the entire structural and functional connectome that can be analyzed by a single GCN. Given prior evidence that the structure-function relationships markedly vary across the neocortex \cite{Rodriguez-Vazquez2019}, we propose learning the weights (quantifying the coupling strength of SC and FC patterns for each region) of the inter-network edges during the end-to-end training. We tested our proposal, called Joint-GCN, on the public data collected by the National Consortium on Alcohol and Neurodevelopment in Adolescence (NCANDA) \cite{Brown2015}. By predicting age and sex of the participants from SC matrices (extracted from DTI) and FC matrices (extracted from rs-fMRI), our joint graph has a higher  prediction accuracy than unimodal analysis (based on either SC or FC networks alone) and than existing multi-modal approaches that disjoint the SC and FC graphs. The learned weights of inter-network edges highlight the non-uniform coupling strength between SC and FC over the human cortex.

\section{Method}

\begin{figure}[!t]
    \centering
    \includegraphics[trim=0 90 0 105, clip,width=1\linewidth]{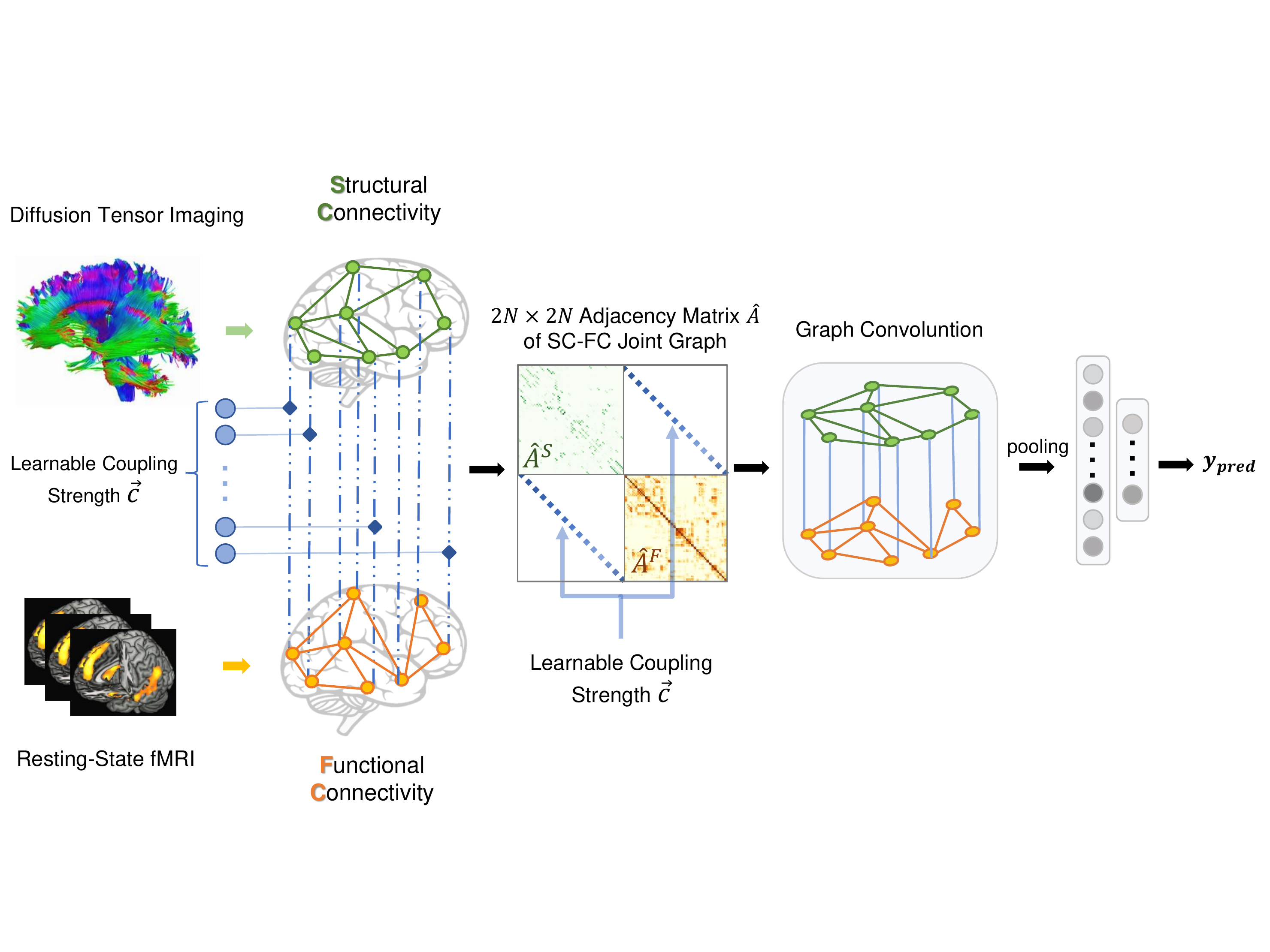}
    \caption{Proposed Joint-GCN Method: Structural and functional graphs are extracted from DTI and rs-fMRI scans respectively. The two networks are then joined via inter-network edges, whose edge weights are learnable parameters. The adjacency matrix of the joint SC-FC graph is then analyzed by a single graph convolutional network.}
    \label{fig:method}
\end{figure}

We first review the traditional graph convolution defined on a single network (SC or FC). We then describe the construction of the joint SC-FC network and model the SC-FC coupling as learnable weights of inter-network edges. Fig. \ref{fig:method} provides an overview of the proposed Joint-GCN.\\
~\\
\textbf{Graph Convolution for Individual Graphs}. We assume a brain is parcellated into $N$ regions of interest (ROIs), so that a tractography procedure (using a DTI scan) quantifies the strength of the SC between all pairs of ROIs. This SC network is characterized by a graph
$\mathcal{G}^{S}(\mathcal{V}^S,\mathcal{E}^S)$ with $\mathcal{V}^S$ being the node set ($N$ ROIs) and $\mathcal{E}^S$ being the edge set. Edge weights between ROI pairs are encoded by an adjacency matrix $A^S \in \mathbb{R}^{N\times N}$. Let $X^S\in \mathbb{R}^{N\times M}$ be $M\text{-dimensional}$ node features defined for the $N$ ROIs, and $\hat A^S=D^{-\frac{1}{2}}A^SD^{-\frac{1}{2}}$ be the normalized adjacency matrix, where $D$ is the degree matrix with $D_{ii}=\sum_{j} A^S_{ij}$. A typical graph convolution $f$ is
\begin{equation}
    f(X^S; A^S) = \sigma(\hat A^S X^S W^S),
\end{equation}
where $\sigma(\cdot)$ is a non-linear activation function and $W^S$ is the convolutional weights to be learned. 

On the other hand, we quantify the strength of FC between a pair of ROIs by correlating their BOLD signals recorded in the rs-fMRI scan. As such, the FC network is characterized by another graph $\mathcal{G}^F(\mathcal{V}^F,\mathcal{E}^F)$, and the corresponding graph convolution is defined with respect to the normalized FC adjacency matrix being $\hat A^F$. Note, $\mathcal{V}^{S}$ and $\mathcal{V}^{F}$ represent the same set of ROIs, whereas $\hat A^S$ and $\hat A^F$ are generally different, reflecting the divergence between SC and FC patterns. Given the SC and FC graphs of a subject, our goal is to use a GCN model to predict the target label (e.g., age or diagnosis) of the subject. \\ 

\noindent\textbf{Joint SC-FC Graph}. Instead of using separate GCNs to analyze the two graphs, we merge $\mathcal{G}^{S}$ and $\mathcal{G}^{F}$ into a joint graph $\mathcal{G}=\{\mathcal{V},\mathcal{E}\}$ encoding the entire SC-FC connectome profile. The joint node set $\mathcal{V}=\{\mathcal{V}^S,\mathcal{V}^F\}$ has a size of $2N$, i.e., the duplication of the $N$ ROIs. We then add $N$ inter-network edges $\mathcal{E}^C$ to connect corresponding nodes in $\mathcal{V}^S$ and $\mathcal{V}^F$. The joint edge set is the combination of $\mathcal{E}=\{\mathcal{E}^S,\mathcal{E}^F,\mathcal{E}^C\}$ (Fig. \ref{fig:method}). Using this combination, the SC and FC profiles associated with the same brain region are coupled together so that the joint SC-FC graph is analyzed by a single GCN.

~\\
\textbf{Learnable SC-FC Coupling Strength}. Recent studies suggest the strength of
SC-FC coupling is not uniform across the brain but exhibits variation in parallel to certain cortical hierarchies \cite{baum2019,Rodriguez-Vazquez2019}. To support this hypothesis, we set the weights of the $N$ inter-network edges to be learnable parameters. Let $\vec c=\{c_1,...,c_N\}$ be a row vector encoding the coupling strength of $\mathcal{E}^C$ with each $c_i \in [0,1]$. We then construct the joint $2N \times 2N$ normalized adjacency matrix of the SC-FC joint graph as (Fig. \ref{fig:method}) 
\begin{equation}
    \hat A=\begin{bmatrix} \hat A^S_{} &  \text{diag}(\vec c)\\ \text{diag}(\vec c) & \hat A^F_{} \end{bmatrix},
\end{equation}
where diag($\cdot$) denotes the diagonalization of a vector. Let $X=\begin{bmatrix} X^S\\ X^F\end{bmatrix}$ denote the vertical concatenation between structural and functional node features. The graph convolution for the joint graph is defined by 
\begin{equation}\label{eq:joint_gc}
    f(X,A) = \sigma(\hat A X W) = \sigma(\begin{bmatrix}\hat A^{S}X^S+\vec c \otimes {X^F}\\{\hat A^F X^F+\vec c \otimes X^S}\end{bmatrix}W),
.\end{equation} 
where $W$ defines the convolution weights for the joint graph and $\otimes$ is element-wise multiplication. Equation \ref{eq:joint_gc} suggests that through our joint graph convolution, the functional information encoded in $X^F$ is propagated into the structural network through $\mathcal{E}^C$ and the amount of propagation aligns with the SC-FC coupling strength encoded in $\vec c$. Vice versa, structural features can also propagate into the functional network through the convolution.\\ 

\noindent\textbf{Joint-GCN}. To embed the learnable $\vec c$ into the end-to-end network, we construct a ``dummy" network that contains a single fully connected layer applied to a dummy input scalar. With a sigmoid activation, the $N$-dimensional output of the dummy network is then viewed as $\vec c$ and enters the diagonal entries in the off-diagonal blocks of the joint normalized adjacency matrix. We simply use the sum of the node degree and centrality within the SC and FC networks as node features, which are then fed into  
a standard GCN with the aforementioned adjacency matrix. The GCN contains one layer of graph convolution with input dimension 80 and output dimension 40 followed by SELU \cite{klambauer2017self} activation, mean pooling, batch-normalization, and two fully connected layers with dimension 1600 and 128 respectively (Fig. \ref{fig:method}). 

\section{Experimental Settings}
Using machine learning approaches to predict age and sex of youths based on their neuroimaging data is becoming a popular approach for understanding the dynamic neurodevelopment and the emerging sexual differences during adolescence. We illustrate the potential of Joint-GCN for these prediction tasks based on the brain connectome data provided by the NCANDA study \cite{Brown2015}.\\

\noindent\textbf{Data}. We used longitudinal rs-fMRI and DTI data from 662 NCANDA participants (328 males and 334 females), who were 12 to 21 years old at their baseline visits and were scanned annually for up to 7 years. Data were acquired on GE and Siemens scanners. Scanner type did not significantly ($p>$0.05) correlate with age (two-sample \textit{t}-test) and sex (Fisher's exact test). For each participant, we selected visits where the participant had no-to-low alcohol intake over the past year according to the Cahalan scale \cite{Zhao2020}. This selection simultaneously resulted in 1976 pairs of rs-fMRI and DTI scans (3 visits per participant on average).\\

\noindent\textbf{Preprocessing}. rs-fMRI data were preprocessed using the publicly available NCANDA pipeline \cite{zhao2019hbm}, which consists of motion correction, outlier‐detection, detrending, physiological noise removal, and both temporal (low pass frequency: 0.1, high pass frequency: 0.01) and spatial smoothing. We then computed the mean BOLD signal within 80 cortical regions by first aligning the mean BOLD image to the subject-specific T1‐weighted MRI and then non-rigidly registering the T1 MRI to the SRI24 atlas \cite{SRI24}. Finally, an $80\times80$ FC connectivity matrix was constructed as the Fisher-transformed Pearson correlation of the BOLD signal between pairwise ROIs. Negative connectivities were set to zero to remove potentially artificial anti-correlation introduced in the rsfMRI preprocessing \cite{weissenbacher2009}.

The public NCANDA DTI preprocessing pipeline included skull stripping, bad single shots removal, and both echo-planar and Eddy-current distortion correction \cite{Zhao2020}. The whole-brain boundary of gray-matter white-matter was extracted by the procedure described in \cite{baum2019} and was then parcellated into the same set of 80 ROIs defined in the rs-fMRI preprocessing procedure. Probabilistic tractography was performed by FSL bedpostx and FSL probtrackx, which initiated 10,000 streamlines from each boundary ROI. Finally, each entry in the $80\times80$ SC connectivity matrix recorded the log of the number of probabilistic streamlines connecting a pair of brain regions.\\
~\\
\noindent\textbf{Model Implementation}. We implemented Joint-GCN in Pytorch 1.10 and trained the model using the SGD optimizer for 1000 epochs with a batch size of 64. We used mean squared error as the training loss for age prediction and the binary cross-entropy as the loss for sex prediction after applying a sigmoid activation to the network output. We used a learning rate of 0.01 in the sex prediction task and 0.001 in age prediction. To mitigate differences in the input data associated with the scanner type, a binary variable encoding the scanner type was concatenated with learned features before the last fully connected layer.\\
~\\
\noindent\textbf{Evaluation}. We evaluated our proposed Joint-GCN model on the 1976 pairs of DTI and rs-fMRI by running 5-fold cross-validation for 10 times. The five folds were divided on the subject level to ensure data of each subject belonged to the same fold. Although our approach can be extended to a longitudinal setting, we focused on a cross-sectional analysis so the across-visit dependency within an individual was not considered.
For sex experiments, we chose the classification accuracy (ACC) and Area under the ROC Curve (AUC) scores as evaluation metrics. For age prediction, we evaluated the Mean Absolute Error (MAE) and the Pearson's Correlation Coefficient (PCC) between predicted and ground truth ages. Average and standard deviation of these metrics over the 10 cross-validation runs were reported. Lastly, we derived the coupling strength of the 80 ROIs for either prediction task by averaging the learned $\vec c$ over the five folds and then averaging the corresponding two regions across left and right hemispheres.\\ 

\noindent\textbf{Baselines}. To examine the effectiveness of the our Joint-GCN model, we first compared it with single-modality models that only analyzed one type of network (either FC or SC). In this setting, we chose GCN, Multi Layer Perceptron (MLP), and Support Vector Machine (SVM) as the baselines. The MLP was a 2-layer network with 20 hidden units making predictions based on node features. The SVM used either FC or SC adjacency matrices as input features. Next, we compared our model with several existing multi-modal GCN approaches that took both SC and FC networks as inputs. We first implemented a multi-modal version of MLP, which concatenated outputs from the two modality-specific MLPs before entering another fully connected layer of dimension 40. The second approach \cite{DSouza} used a Matrix Auto-Encoder (Matrix-AE) to map FC to SC resulting in a low dimensional manifold embedding that can be used for classification. The third approach \cite{Jing2020}, UBNfs, used a multi-kernel method to fuse FC and SC to produce a unified brain network and used a standard SVM for classification. Next, we implemented Multi-View GCN (MV-GCN) proposed in \cite{zhang2018multi}, which applied two separate GCNs to extract features from SC and FC networks and then used the merged feature for the final prediction. Lastly, we tested SCP-GCN \cite{Liu2019} (using their code) that defined adjacency matrix with respect to the SC graph and treated the FC matrix as node features. To align with our Joint-GCN implementation, scanner type was concatenated to the features before the final prediction in all baselines. 


\section{Results}
\begin{figure}[!t]
    \centering
    \subfloat[][Age prediction]{\includegraphics[trim=0 0 0 0, clip,width=0.33\linewidth]{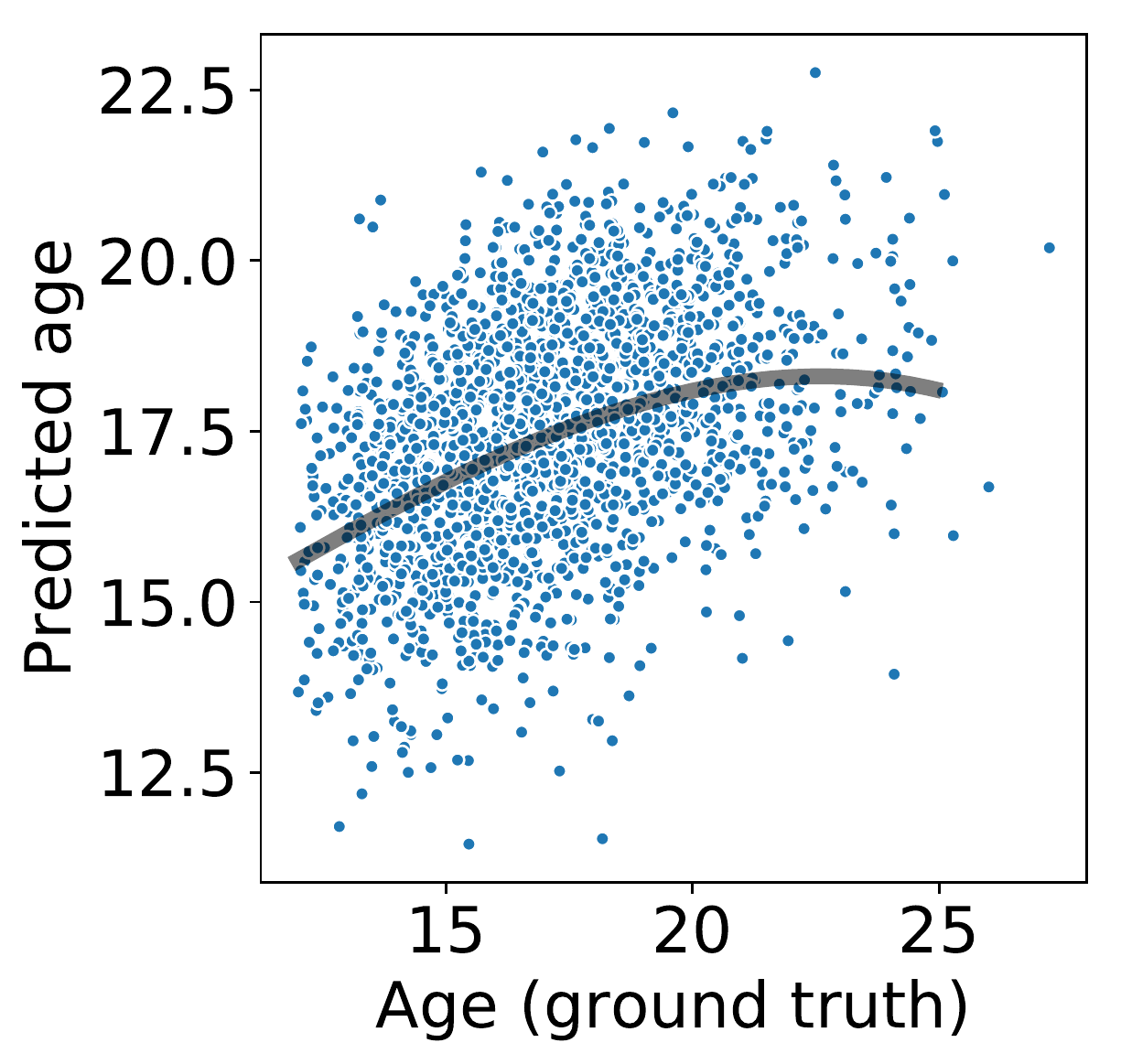}}
    \subfloat[][Sex classification]{\includegraphics[trim=0 -25 0 0, clip,width=0.3\linewidth]{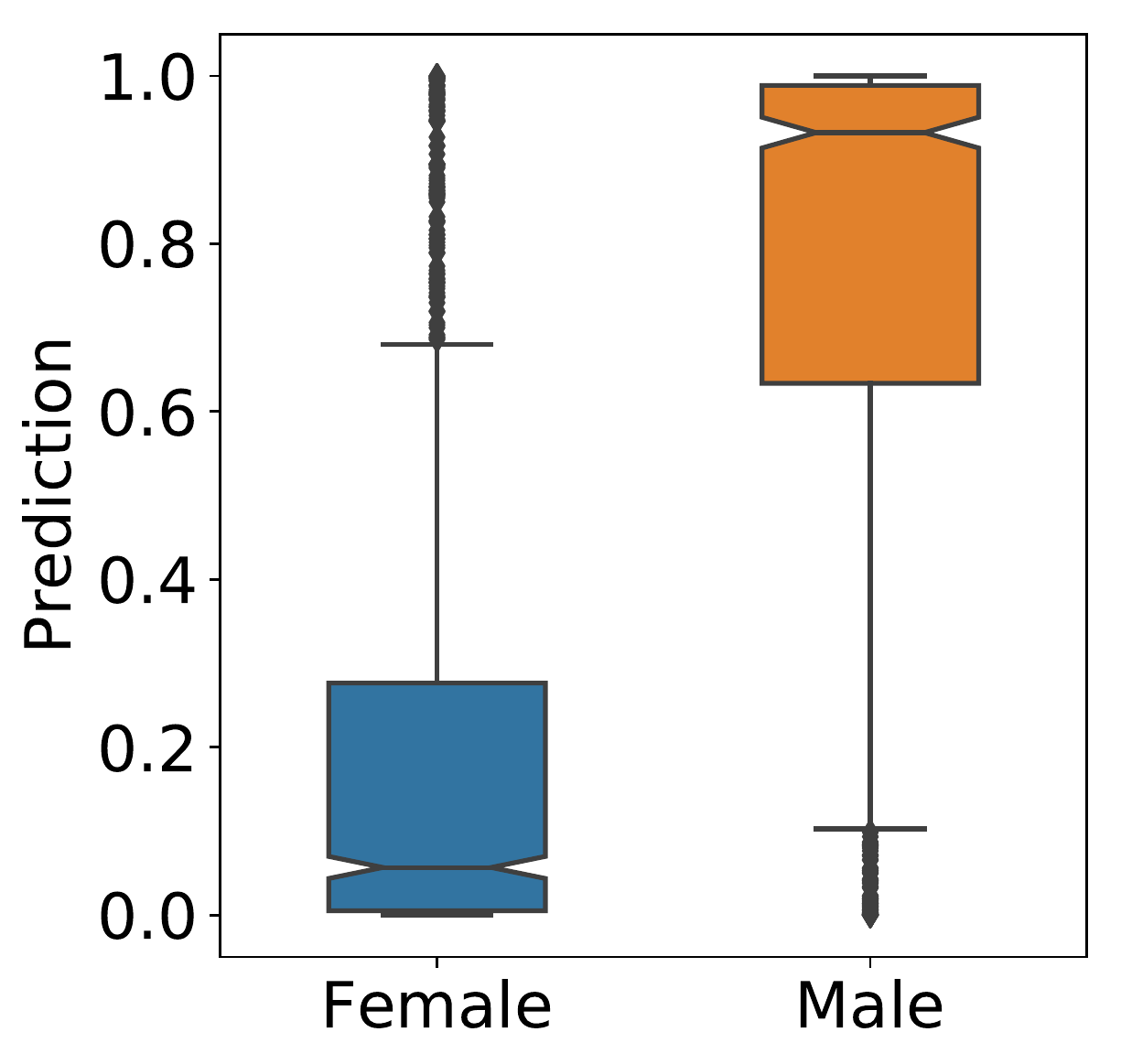}}
    \subfloat[][SC-FC coupling strength]{\includegraphics[width=0.33\linewidth]{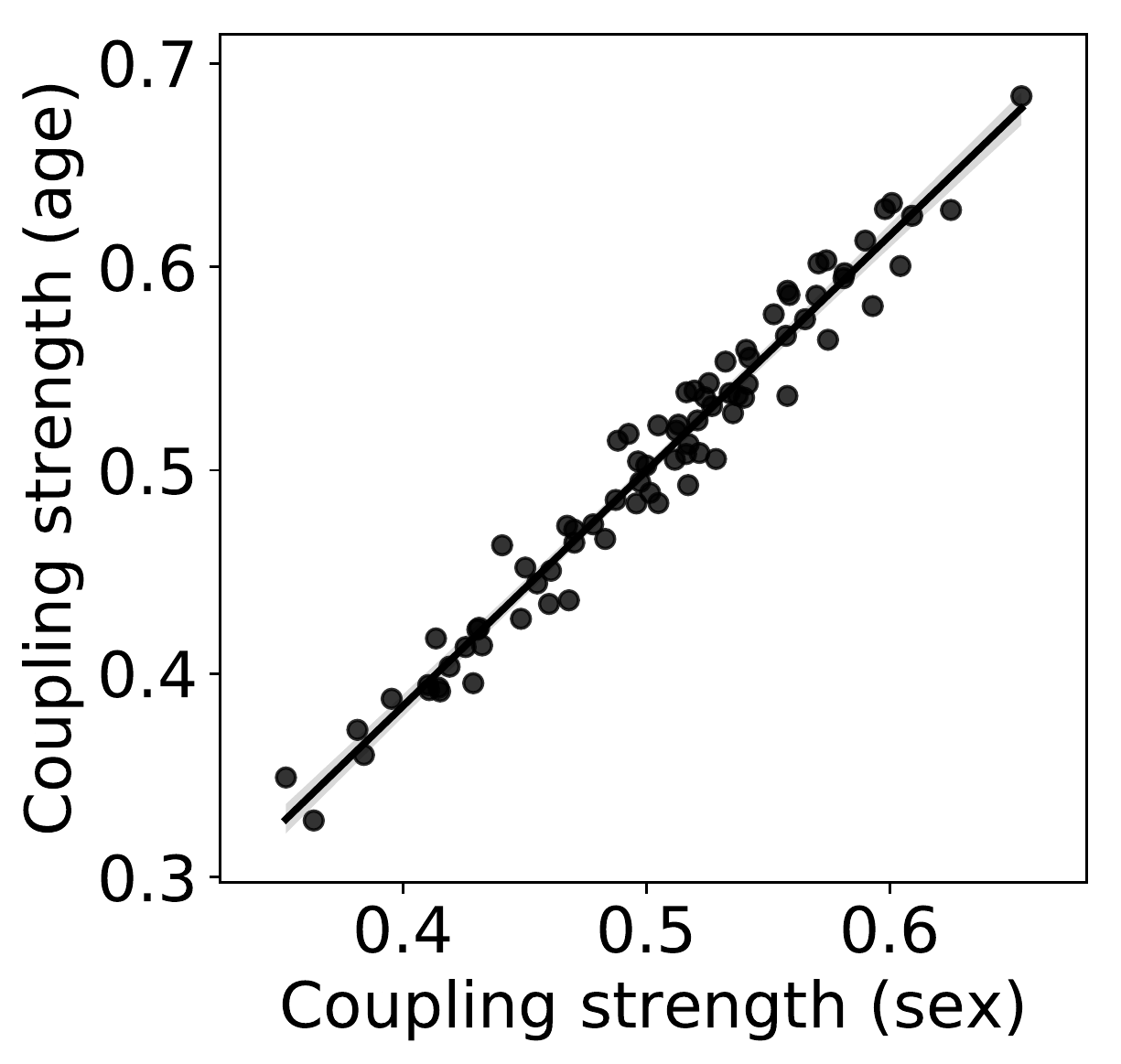}}\\
    \subfloat[][Regional coupling strength displayed on the cortex]{\includegraphics[width=1\linewidth]{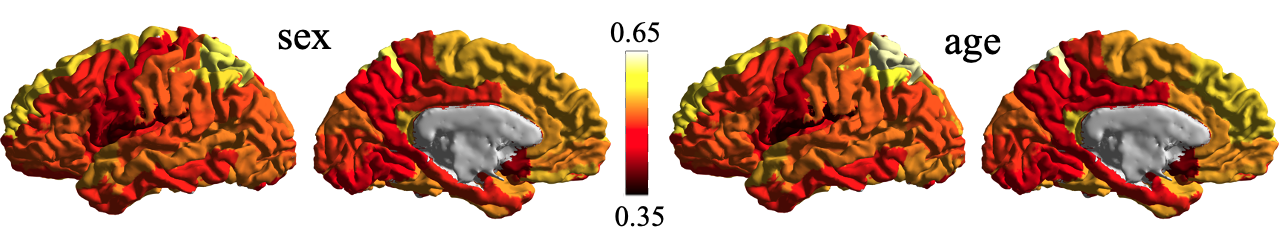}}
    \caption{Predicted values of (a) age and (b) sex for the NCANDA participants by Joint-GCN; (c) The ROI-specific weights quantifying the coupling strength between SC-FC learned in the sex prediction task significantly correlated with the weights learned for age prediction ($p<0.01$, Pearson's $r$); (d) The learned coupling strength displayed on the brain cortex. }
    \label{fig:coupling}
\end{figure}

\noindent\textbf{Predictions by Joint-GCN}. Based on 10 runs of 5-fold cross validation, Joint-GCN resulted in an average of 84.9\% accuracy for sex classification and an MAE of 1.95 years for age prediction. Fig. \ref{fig:coupling}(a)(b) shows the predicted values of the NCANDA participants by our model. We observe that the predicted age non-linearly correlated with ground truth; that is, our model could only successfully stratify the age differences at younger ages. Notably, the MAE for the younger cohort (MAE=1.86, age $<$ 18 years) was significantly lower ($p<0.01$, two-sample $t$-test) than that for the older cohort (MAE=2.12, age $\geq$ 18 years). This result comports with adolescent neurodevelopment being more pronounced in the younger age but generally slows upon early adulthood \cite{pujol1993does}. The cubic fitting between predicted age and ground truth in Fig. \ref{fig:coupling}(a) was also inline with the ``inverted U-shape" of the developmental trajectory of functional and structural connectome during adolescence \cite{Zhao2020}.\\

\noindent\textbf{Comparison with Baselines}. Table \ref{tab:result} shows the accuracy of sex classification and age prediction for all the comparison methods. GCN-based (including ours) approaches generally resulted in higher accuracy than the others (i.e. MLP and SVM). 
The higher accuracy of GCN-based methods shows the efficacy of graph convolution in extracting informative features from network data. Compared to the single GCN applied to FC alone and SC alone, multi-view GCN and our Joint-GCN resulted in more accurate sex classification as well as higher PCC in age prediction. This phenomenon indicates SC and FC networks contain complementary information about the brain connectome that is useful for identifying sex and age differences. 
Notably, if we modeled the coupling by adding inter-network edges between every pair of ROIs (instead of using ROI-specific edges), the prediction accuracy was significantly lowered (sex ACC: 77\%, age PCC: 0.34).
Lastly, our proposed Joint-GCN resulted in the highest scores in three metrics and the second best in MAE. The improvement in PCC over the second best approach (MV-GCN) was statistically significant ($p<0.01$, two-sample $t$-test), and so was the improvement in ACC for sex prediction ($p=0.015$, McNemar's test).
\begin{table}[!t]
\centering
\caption{\label{tab:result}Accuracy of all comparison models trained on single modality or multi-modal brain connectome for predicting sex or
age labels of the NCANDA participants. Results were averaged across the 5 folds for 10 times of random dataset splits. Highest accuracy is typeset in bold.}
\begin{tabular}{>{\raggedright}p{0.15\textwidth} >{\raggedright }p{0.21\textwidth} | >{\centering}p{0.13\textwidth} >{\centering}p{0.16\textwidth} | >{\centering}p{0.13\textwidth}  >{\centering\arraybackslash}p{0.17\textwidth}} 
\hline
{Modality}                   &Method &\multicolumn{2}{c}{Sex Classification}    &\multicolumn{2}{c}{Age Prediction}\\\cline{3-6}
                             &       & ACC(\%) $\uparrow$   & AUC $\uparrow$          & MAE $\downarrow$         & PCC$\uparrow$\\     \hline
{FC}                         &MLP   & 60.0$\pm$0.1 & 0.612$\pm$0.016    
                             &2.03$\pm$0.006  & 0.117$\pm$0.018  \\
                             &GCN\cite{kipf2016semi}      & 80.9$\pm$0.6  & 0.870$\pm$0.007   & 2.08$\pm$0.023   & 0.316$\pm$0.018 \\ 
                             &SVM      &76.2$\pm$0.6                  &0.838$\pm$0.005          &2.72$\pm$0.042          &0.230$\pm$0.015     \\ \hline
{SC}                         & MLP       & 57.8$\pm$2.0  & 0.598$\pm$0.032                              & 2.03$\pm$0.005 &0.154$\pm$0.009  \\
                             & GCN\cite{kipf2016semi}                     & 78.4$\pm$0.6          & 0.838$\pm$0.009        & 1.98$\pm$0.026          & 0.315$\pm$0.015\\
                             & SVM                     &74.1$\pm$1.0          & 0.816$\pm$0.010        & 2.11$\pm$0.027          & 0.316$\pm$0.016\\ \hline
{Multi-modal}                & MLP     &   61.9$\pm$1.5  
                             & 0.647$\pm$0.018  & 2.03$\pm$0.007  
                             & 0.145$\pm$0.014\\
                             & Matrix-AE\cite{DSouza} &   68.8$\pm$3.4
                             & 0.772$\pm$0.031   & 2.09$\pm$0.036   &0.246$\pm$0.019  \\
                             & UBNfs\cite{Jing2020} & 80.4$\pm$0.5  
                             & 0.884$\pm$0.003          & \textbf{1.93$\pm$0.009}   & 0.304$\pm$0.009 \\
                             & SCP-GCN\cite{Liu2019} & 70.8$\pm$0.7        & 0.759$\pm$0.011         & 1.96$\pm$0.023   &0.303$\pm$0.014\\
                             &MV-GCN\cite{zhang2018multi} & 83.1$\pm$0.7     & 0.893$\pm$0.005       & 2.03$\pm$0.034    & 0.351$\pm$0.017          \\
                            &{Joint-GCN(Ours)}               & \textbf{84.9$\pm$0.6}          & \textbf{0.907$\pm$0.005}          
                            &1.95$\pm$0.006
                            &\textbf{0.386$\pm$0.008}      \\\hline
\end{tabular}\\
\end{table}

~\\
\noindent\textbf{Learned SC-FC Coupling Strength}. Fig. \ref{fig:coupling}(d) displays the coupling strength of the 40 bilateral brain regions learned in the age and sex prediction tasks. We observe that learned coupling strength highly coincided between the two prediction tasks despite the training of the two models being independent. This indicates our learning framework can extract intrinsic  SC-FC relationships in the adolescent brain, which contributes to the identification of age and sex differences.  Supported by prior studies  \cite{baum2019,Rodriguez-Vazquez2019}, the learned coupling strength was not uniform across the neocortex but varied from 0.35 to 0.65 (0.5 $\pm$ 0.05). The top 3 regions with the strongest SC-FC coupling are the superior and the inferior parietal lobe and superior frontal lobe. The lowest coupling strength was measured for the rolandic operculum, olfactory cortex, and frontal inferior operculum. As recent findings have identified macroscopic spatial gradients as the primary organizing principle of brain networks \cite{baum2019,Rodriguez-Vazquez2019}, we computed the principal gradient of functional connectivity for each ROI \cite{Rodriguez-Vazquez2019}. These gradients negatively correlated with the learned coupling strength on a trend level ($p$=0.075) with unimodal regions generally having stronger and transmodal regions weaker coupling strength. We hypothesize that the trend-level correlation would become more pronounced once we replace our 80-ROI parcellation with a finer one as in \cite{baum2019,Rodriguez-Vazquez2019}. Nevertheless, the above result points to the potential of Joint-GCN in learning useful structural-functional properties of the brain.

\section{Conclusion and Future Work}
In this paper, we proposed a novel framework, called Joint-GCN, for analyzing brain connectome quantified by multi-modal neuroimaging data from DTI and rs-fMRI. Extending prior studies on applying GCN to structural and functional graphs, our work underscored the importance of modeling the coupling between structural and functional connectivity for prediction tasks based on brain connectome. Based on the adolescent neuroimaging dataset provided by NCANDA, we showed the potential of our framework to accurately characterize the protracted development of structural and functional brain connectivity and the emerging sex differences during youth. One limitation of our study is that we did not focus on modeling the across-visit relationships within the longitudinal data of NCANDA participants (an orthogonal research direction). Rather, we only considered network metrics (degree and centrality) as node features to focus on exploring the model's capability in analyzing network topology. Nevertheless, our model presents a useful data-driven approach to model the complex hierarchical neural systems that have broad relevance for healthy aging and abnormalities associated with neuropsychological diseases.

\textbf{Acknowledgment.} This  research  was  supported  in  part  by  NIH U24 AA021697, K99 AA028840, and Stanford HAI GCP Credit. The data were part of the public NCANDA data releases NCANDA\_PUBLIC\_6Y\_REDCAP\_V04 \cite{ncanda_redcap_release}, NCANDA\_PUBLIC\_6Y\_DIFFUSION\_V01 \cite{ncanda_diffusion_release}, and NCANDA\_PUBLIC\_6Y\\\_RESTINGSTATE\_V01   \cite{ncanda_resting_release}, whose collection and distribution were supported by NIH funding AA021697, AA021695, AA021692, AA021696, AA021681, AA021690, and AA02169.
%
%
%
\bibliographystyle{splncs04}
\bibliography{miccai.bib}
\end{document}